\documentclass[conference,compsoc, times]{IEEEtran}

\usepackage{cite}
\usepackage{amsmath,amssymb,amsfonts,amsthm}
\usepackage{algorithmic}
\usepackage{graphicx}
\usepackage{textcomp}
\usepackage{xcolor}
\usepackage{booktabs}
\usepackage[hidelinks]{hyperref}

\usepackage{url}
\usepackage[draft]{todonotes}
\usepackage{xspace}
\usepackage{pifont} 
\usepackage{balance}

\numberwithin{equation}{section}
\theoremstyle{definition}

\begin{document}

\title{SoK: Log Based Transparency Enhancing Technologies}

\author{
\IEEEauthorblockN{Alexander Hicks}
\IEEEauthorblockA{University College London}
}

\maketitle

\begin{abstract}
This paper systematizes log based Transparency Enhancing Technologies.
Based on work on transparency from multiple disciplines we outline the purpose, usefulness, and pitfalls of transparency.

We describe the mechanisms that allow log based transparency enhancing technologies to be implemented, in particular logging mechanisms, sanitisation mechanisms and the trade-offs with privacy, data release and query mechanisms, and how transparency relates to the external mechanisms that enable contesting a system and holding system operators accountable.

We illustrate this with two examples, Certificate Transparency and cryptocurrencies, and show the role that transparency plays in their function as well as the issues these systems face in delivering transparency.
\end{abstract}

\section{Introduction}\label{sec:intro}
As systems perform operations and assist decisions that can have an important impact on a person's life, transparency is often suggested as a way of identifying flaws in a system, enabling accountability and making it more likely that flaws are rectified and their impacts mitigated.

Transparency, however, is a complex property to require from a system.
It does not entail any specific meaning or way of implementing transparency, particularly in systems deployed in an environment that is adversarial to the accountability that transparency should enable.
What information is revealed? In what form? By who? To whom? How?
As a result, transparency does not always work as desired and is sometimes even counterproductive~\cite{taylor2016transparency}.

In this paper, we consider achieving transparency based on logging mechanisms.
This involves technical considerations, such as logging, sanitising, releasing and querying data, as well as non-technical external mechanisms that determine what can be done once transparency is in place.
Our aim is to provide a systematisation that brings the relevant aspect of each mechanism into one view of log based transparency enhancing technologies.

\subsubsection*{Outline of the paper}
We motivate applying transparency to computer systems and give an overview of transparency and criticisms of transparency in Section~\ref{sec:overview}, before outlining log based transparency enhancing technologies based on four essential mechanisms in section~\ref{sec:essential}: logging, sanitisation, release and query, and external mechanisms.

In Section~\ref{sec:def} we discuss threats to transparency mapped to the essential mechanisms outlined in the previous section and editorial control and individual evidence.

We consider the infrastructure that supports logging in Section~\ref{sec:infra} and the interaction between transparency and privacy in Section~\ref{sec:privacy}.
To illustrate our discussion we provide in Section~\ref{sec:cases} two case studies of transparency systems, Certificate Transparency and cryptocurrencies.


We discuss related work in Section~\ref{sec:related}, before concluding in Section~\ref{sec:conclusion}.

\subsubsection*{Methodology}
Transparency is a broad topic that many fields have independently studied, not all of which can be covered here.
For work on transparency from other fields, we have, therefore, focused on work from Law, Philosophy, Business, and Economics, which provide a basis for thinking about transparency and computer systems.

Because of our focus on log based transparency enhancing technologies and the security of the mechanisms involved in such systems, we have endeavoured to find relevant papers from the information security literature by going through publications at major conferences like IEEE S\&P, ACM CCS, NDSS, Usenix Security, PETS, and ACM FAccT, as well as searching for papers from other smaller conferences, workshops, and journals, including those in adjacent fields (e.g., HCI, STS).
Work that relates to transparency but not directly to log based transparency enhancing technologies (e.g., work on transparent machine learning) is out of scope and, therefore, not included.

\section{A Short Overview Of Transparency}\label{sec:overview}
Transparency can be defined as ``the quality of being done in an open way without secrets''~\cite{cambridgetransparency}.
Applied to an organisation, it can mean that the organisation is ``open, public; having the property that theories and practices are publicly visible, thereby reducing the chance of corruption''~\cite{wiktionarytransparent}.

These definitions express the basic intuition that if something is being done transparently then it cannot be done badly without it being noticeable.
As Brandeis put it, ``sunlight is said to be the best of disinfectants''~\cite{alma991026159019703276}.

This should create an incentive to ensure that things are done well if there is a high likelihood of being held to account, making transparency an enabler of accountability or other ethical principles (e.g., safety, welfare)~\cite{turilli2009ethics}.

Transparency such as open data practices promoted by both governments~\cite{uk,obama2009transparency} and academics~\cite{robinson2008government,degbelo2018increasing}, lead to the public release of data that is used to determine policy.
Open data practices are also used in scientific research to allow results to be reproduced, further research to be conducted, and new algorithms to be benchmarked.


In a more bottom-up manner, freedom of information laws have enabled the media, NGOs, and the public, to make requests for information that can be used to hold public authorities to account.
Other regulations, such as the GDPR~\cite{recital} give individuals the right to request a copy of their personal data that is held by a controller (Article 15), require that personal data should be ``processed [...] in a transparent manner in relation to individuals'' (Article 5), and as general data processing principles that ``it should be transparent to natural persons that personal data concerning them are collected, used, consulted or otherwise processed and to what extent the personal data are or will be processed'' and ``the principle of transparency requires that any information and communication relating to the processing of those personal data be easily accessible and easy to understand, and that clear and plain language be used'' (Recital 39).


Access to data also helps mitigate information asymmetries.
The work of Akerlof on information asymmetries in markets~\cite{akerlof1978market} has led to security economics re-framing many security issues (e.g., software security) as problems of information asymmetry~\cite{anderson2006economics,anderson2001information}, which can be dealt with by requiring data standards and disclosures.

This ties into Saltzer and Schroeder's open design principle~\cite{saltzer1975protection}.
Most security mechanisms (e.g., cryptographic algorithms) are open, enabling users with the technical knowledge required to assess a system's code or specification to determine whether they want to rely on the system.
Beyond the specification and code of a system, nutrition labels for datasets~\cite{gebru2021datasheets,holland2018dataset,datalabel} and models~\cite{mitchell2019model}, and privacy labels ~\cite{kelley2009nutrition,kelley2010standardizing}, have been proposed.

\subsection{Transparency matters for computer systems}
Security mechanisms are designed to allow certain properties of systems (or the data they operate on) to hold e.g., integrity, confidentiality, or availability, but no system is perfect.
However, designs can be flawed, implementations can suffer from software bugs or faulty hardware, and systems can be misused.
Security Economics tells us that we should not expect perfect security in practice, even when technical mechanisms appear to be sufficient in principle~\cite{anderson2001information,anderson2006economics}.

Even if we perfected the design of systems, designing and implementing complex systems that are entirely formally verified is currently unrealistic and would not prevent harms that occur because of a system that, operating as intended, applies harmful norms~\cite{hicks2022transparency}.
Information is routinely copied, aggregated, and analysed across networks operated by different parties, rendering strict enforcement mechanisms impractical compared to relying on accountability~\cite{weitzner2008information}.
Notions of appropriate use may depend on the data itself as well as the context -- an emergency that requires immediate access to medical data would render any strict security mechanism preventing this useless~\cite{feigenbaum2011accountability}.

More generally, evaluating strict compliance with norms assumes that there are reliable norms, despite many systems operating in grey areas~\cite{hicks2022transparency}.
As systems grow in size, complexity, and scope of applications that impact people's lives, the ability to evaluate systems is increasingly important, not only for auditors or regulators but also for users who may change how they interact with the system~\cite{10.1145/3290605.3300724}.

Evaluating systems is not new, and system operators routinely do so internally but this does not always work to reduce the harm that a faulty system can cause.
There can be issues with how the evaluation is done e.g., flawed mechanisms or metrics.
Even if a system operator detects faults in the system it operates, it still has to address these faults and may not do so if it does not have the incentive or the capacity (technical or economic) to do so.

Systems are not inherently inscrutable~\cite{kroll2018fallacy}, but those to whom harm is caused cannot necessarily detect or show that the system is at fault, despite being those that have a greater incentive to do so.
Access control mechanisms that regulate rights over a system tend to favour those who design or commission these access control mechanisms i.e., system operators, rather than those subject to the system who have no ability to access useful information via the system itself.

Privilege over information about the system, such as known error logs, means that system operators can manipulate disclosure procedures to their advantage~\cite{marshall2020recommendations}.
This includes many types of systems, such as accounting systems(e.g., Horizon, linked to one of the biggest miscarriage of justice in the UK~\cite{payout}), breathalysers (See Bellovin et al.~\cite{bellovin2021seeking}), and newer data processing systems that result in unfair and harmful outcomes~\cite{angwin2016machine,barocas2016big}.


Transparency enhancing technologies offer a way to not only provide trustworthy transparency through the use of security mechanisms but also to scale transparency.
For example, the IPCO, which audits law enforcement requests for telecommunications data in the UK, perform local inspections of a limited amount of offices to produce their audit~\cite{ipco2020}.
Transparency enhancing technologies could allow for larger and more efficient audits of many practices.


Moreover, while transparency can have negative effects on people if they respond to transparency with hiding behaviour, impacting their performance~\cite{bernstein2012transparency}, the opposite could be true for computational systems with secure transparency mechanisms because the performance of such systems is determined by the code and infrastructure it runs on, not on whether or not it is being observed.
Given two systems that perform similarly, if transparency is cheap enough to implement and expensive enough to cheat once implemented (e.g., breaking the logging mechanism's cryptographic properties), the honest transparent system will be cheaper to operate than the one that tries to cheat transparency, which should make it more competitive. (That is unless the system is so broken in the first place that whatever is revealed by transparency condemns the system.)

Transparency can also be seen as a tool for efficiency.
Decentralised systems are often desired because they do not rely on a central party, but centralised systems are typically more efficient to operate
They can also make more sense logistically, for systems that either involve sensitive data that cannot be used in an encrypted form for operational reasons or simply to avoid the burden of coordinating many (sometimes unaligned) parties.
A decentralised transparency enhancing technology, overlaid on top of a centralised system with a trustworthy interface between both, can provide a useful compromise between the inherent efficiency and logistical advantages of the centralised system and the lesser trust required by a decentralised system.


\subsection{Forms Of Transparency}
Transparency can take numerous forms based on the direction in which information flows, the type of information that flows, and when it flows.

Directions of transparency are reminiscent of basic access control models e.g., Bell-LaPadula~\cite{bell1973secure} and BIBA~\cite{biba1977integrity}, which determine in which direction (upwards or downwards) information can be read or written.
Unlike many access control mechanisms, however, transparency requires that information leaves the system and be accessible by users with no privilege over the system, and restricts the write access of privileged users over this information.

Concerning the type of information, there can be information about inputs to a system, processes executed within the system, and outputs of the system, where different levels of transparency (or data granularity) matter.
For example, when revealing the inputs to a system, the ordering of inputs can also be important as the ordering of data used to train a model can affect its performance~\cite{shumailov2021manipulating}.

Timing determines when information is made available.
It is uncommon to have real-time transparency when humans are involved as knowingly being surveilled can affect behaviour~\cite{bernstein2014transparency}.
A computer cannot be aware that its actions are being logged but a human user of the computer will be, so this can still be a concern in some cases.
Even for entirely computational systems, transparency may only be useful if there is enough information to obtain an aggregate view of the system's performance but systems such as cryptocurrencies offer a live transparent view of the system.

\subsection{Criticisms of Transparency}

\subsubsection*{Lack of effectiveness}
The assumption that underpins much of the belief in transparency is that it will lead to accountability, better behaviour, and increased public trust.
Criticism of this assumption is centred around the gap between the dissemination of information and its usefulness in enabling sanctions on a misbehaving party~\cite{fox2007uncertain}.

Etzioni has argued that there is little evidence that supports the view that transparency is an effective accountability mechanism~\cite{doi:10.1111/j.1467-9760.2010.00366.x}.
The argument is that transparency is no alternative to regulation (it can only be complementary) because regulations cannot be replaced by offloading the responsibility of demanding and analysing data to citizens without the time or other resources to handle these tasks.

This is backed up by Ferry and Eckersley, who found that, in the UK, the replacement of formal audits with requirements for English local authorities to publish datasets (with little contextual information) weakened accountability~\cite{ferry2015accountability}.
In countries without regulations that implement effective accountability, however, transparency can be effective at bypassing corrupt official audit processes~\cite{ferry2015accountability}.

The issue is that information being transmitted about a bad outcome does not prevent it.
Moreover, it does not prevent future bad outcomes either as it does not, by itself, mitigate their possibility.
A practical example of this is mandated disclosures such as nutrition labels, which do not prevent any nutritional harms that, in any case, are linked to many factors beyond the nutritional value of a food item.
The same is likely to be true with proposals for data and privacy nutrition labels.
A label stating that a dataset has flaws does not prevent anyone from using the dataset and producing a flawed model trained on that dataset.

Research on the effectiveness of privacy labels has also shown that issues of judgement and misdirection could render transparency ineffective~\cite{adjerid2013sleights,acquisti2013gone}.
Developers themselves are not always well equipped to evaluate the labels they create, because privacy is not necessarily their expertise and they may not account for harms that are unknown to them~\cite{li2022understanding}.
If any harm is perceived as originating from the use of a problematic dataset or privacy-invasive system, a system operator will not be prevented from deploying such a system and may also rely on nutritional labels as cover if the process that produces these labels can be influenced.

Yu and Robinson have a similar view on open government technology and data, arguing that while it may allow the public to contribute in new ways, it does not create any government accountability~\cite{yu2011new}.
Open government initiatives generally do not imply any effect on how government works (other than publishing data) so any faulty process is likely to remain in place. 
Thus, open data and transparency may be used as a trojan horse for other political goals~\cite{doi:10.1177/2053951715621568}.

If transparency by itself does not entail accountability, it follows that it also does not necessarily create trust.
Despite greater access to information, for example in the case of government transparency and freedom of information, trust has not increased~\cite{o2002question,o2006transparency,worthy2010more,mabillard2016transparency}.
If transparency only reveals systemic faults, why trust such a system?

\subsubsection*{Restricted transparency}
Obtaining information that is theoretically available, for example through Freedom of Information requests, can also be an issue that requires people to develop specific expertise.
In other cases, the release of bulks of information may also obfuscate important information~\cite{stohl2016digital}.
Even if a party is honest, the release of information implied by transparency does not necessarily imply the effective communication and understanding of that information~\cite{o2006transparency} or that the information that is released is not chosen purposefully to serve a chosen narrative~\cite{adelberg1979narrative}.

These criticisms extend to algorithmic transparency for black box computational systems~\cite{ananny2018seeing,weller2019transparency}.
Burrell distinguishes three forms of opacity in the context of algorithmic systems, opacity as intentional corporate or state secrecy, opacity as technical illiteracy, and opacity as the way algorithms operate at the scale of application~\cite{burrell2016machine}.

Rights such as data subject access requests may also not work well in practice~\cite{ausloos2018shattering}.
This highlights the gap between transparency and other properties e.g., fairness and explainability, of a computational system.
Knowing the inputs, rules, and outcomes of a complex system may not be enough to understand its processes.
Thus, while auditing is necessary and possible, auditing decisions that result from algorithms can still pose a significant challenge~\cite{mittelstadt2016automation}.

Even systems that are open source are not necessarily more or less secure than closed systems~\cite{anderson2005open,schryen2011open} because there are many steps in between code being released in open source form and bugs in the code being identified and fixed, such as having the necessary resources and processes to fix bugs.
Again, this highlights the gap between the availability of information and actions taken based on that information -- in this case, auditing for and fixing vulnerabilities.

\subsubsection*{Tension with privacy and confidentiality}
Another criticism of transparency is that it can cause harm privacy or negatively affect businesses that rely on confidential components in their systems.
This is particularly important for systems that process sensitive data, despite the fact that greater transparency about the sharing and processing of sensitive data may be desirable.

The potential privacy harms brought on by the release of information are also used to restrict transparency.
Freedom of information requests may be refused if they involve the release of personal information that would contravene data protection principles~\cite[Chapter 36, Part II, Section 40]{foiact}.

Similar situations occur when it comes to challenging systems.
For example, Uber invoked privacy concerns to impede a challenge by Uber drivers seeking to obtain information about the system that they were subject to~\cite{uberprivacy}.
More generally, unless compelled to, companies are often extremely reluctant to disclose anything that they can argue falls under commercial confidentiality.

\section{Essential Mechanisms}\label{sec:essential}
\begin{figure}[t]
\centerline{\includegraphics[width=0.9\linewidth]{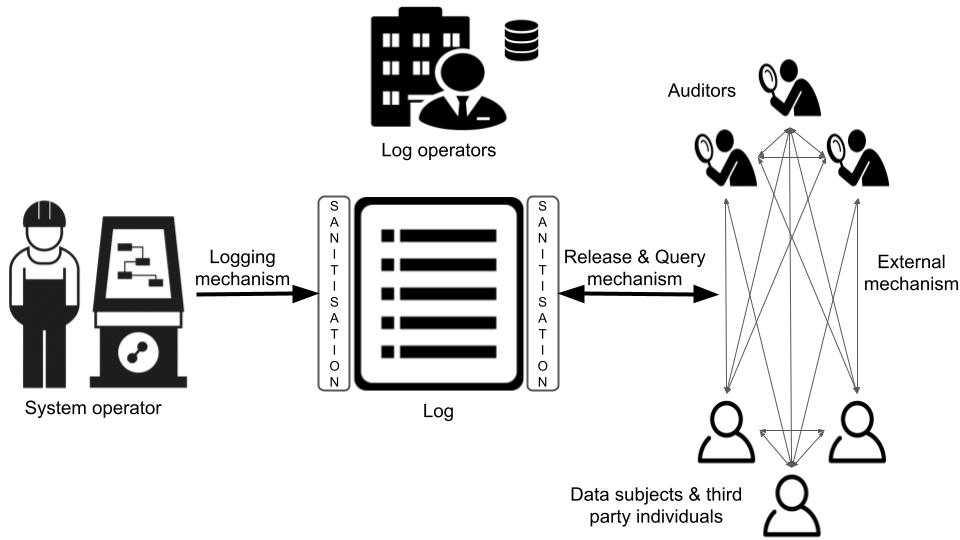}}
  \caption{Summary of essential mechanisms for transparency enhancing technologies (logging, sanitisation, release and query, external) and their place in a transparency process.}
  \label{fig:mechanisms}
\end{figure}

This section introduces transparency enhancing technologies based on logging mechanisms, sanitisation mechanisms that process the data into a format suitable for release, release and query mechanisms, and external mechanisms to make use of transparency.
Figure~\ref{fig:mechanisms} illustrates where each mechanism takes place and the parties it relates to.

Logging involves the system operator of the subject system and log, which is maintained by log operators.

Sanitisation takes place either between the logging mechanism recording information and committing it to the log (e.g., to protect commercially confidential information that even trusted auditors may not see) or before the release and query mechanism (e.g., to allow for both privacy-preserving releases of information and access to raw data depending on the party information is released to, and enforce access control to information).

The release and query mechanism relates the log to the users of transparency i.e., auditors, data subjects, and other third-party individuals, who then relate to each other and take action through external mechanisms.

\subsection{Logging mechanism}

Transparency requires information to be recorded and traceable~\cite{kroll2021outlining}, for example in the form of a chronological list of events or actions that have taken place, a record of the data used by the system to operate, or even a complete record of any byte in a current or past state~\cite{devecsery2014eidetic}.

Secure logging mechanisms have been of interest to cryptographers for a long time ~\cite{bellare1997forward,schneier1998cryptographic,schneier1999secure,chong2003secure,holt2005logcrypt,waters2004building,ma2009new,pulls2013distributed}.
For the purpose of transparency, they have coalesced under authenticated data structures~\cite{tamassia2003authenticated,miller2014authenticated} and transparency overlays~\cite{chase2016transparency}, which are designed to broadly ensure that the log is verifiably append-only, can be used to lookup information, and is consistent i.e., shows the same information to everyone and does not equivocate.
This is typically achieved with Merkle trees or blockchains, although more recent work has also explored the use of append-only dictionaries.

\subsubsection*{Merkle Trees}
Merkle Trees are binary trees based on a hash function $h$ such that each node $i$ takes the value $h_i=h(h_{left(i)}\vert h_{right(i)})$ based on its left and right children.
Given that $h$ is collision-resistant, tamper resistance is guaranteed as modifying any node will result in a different root hash.
This makes it possible to verify the integrity of any data encoded as a leaf in the tree.

A history tree, following the work of Crosby and Wallach~\cite{crosby2009efficient}, grows from left to right and is used by systems like Certificate Transparency~\cite{laurie2014certificate}.
This allows for logarithmic-sized proofs that the log is append-only as new values (e.g., the hashes of new certificates in Certificate Transparency) are added to the log by a log server.
This addition results in a new Merkle tree and root hash, which is signed by the log server.
Because the tree grows from left to right, it is then possible to efficiently verify that the new Merkle tree includes everything that was included in the old one, showing that it is append-only.
Looking up specific certificates, however, requires linear-sized proofs.

As shown by Chase and Meiklejohn~\cite{chase2016transparency} the Certificate Transparency log satisfies \textit{consistency} i.e., a potentially dishonest log server cannot get away with presenting inconsistent versions of the log to different parties, \textit{non-frameability} i.e., parties cannot blame the log server for misbehaviour if it has behaved honestly, and \textit{accountability} i.e., evidence can be used to implicate log servers that promised to include events but then did not. 

A prefix tree, as used by CONIKS~\cite{melara2015coniks} to allow users reliant on a PKI (e.g., for communication apps) to verify the consistency of the public keys of other users, has leaf nodes ordered in lexicographic order.
This makes it efficient to look up values in the tree, although showing that the log is append-only now requires linear-sized proofs.
For example, a client can register name-to-key bindings in the Merkle tree's leaf nodes, which other clients can then lookup on behalf of other users.
To verify a name-to-key binding in the tree, a client checks the signed tree root (STR), which includes the root hash and a hash of the previous STR (successive STRs form a chain), and the inclusion of that name-to-key binding with the path from the root to the leaf node for that name-to-key binding.

Non-inclusion of a name-to-key binding can also be checked by verifying that given an index (i.e., a name), there is no key data mapped to it.

To prevent incidents, clients monitor their user's key bindings do not change unexpectedly and verify that the PKI's identity providers are presenting consistent versions of their key directories to all participants by checking that a provider has correctly signed the STR and that the hash of the previous STR matches what was previously seen.

\subsubsection*{Combining Merkle trees}
A prefix tree and a history tree can be combined to form a verifiable log-backed map~\cite{trillianvds,trillian,keytransparency,ryan2014enhanced}.
(The prefix tree can alternatively be a hash treap~\cite{pulls2015balloon,peeters2016insynd}.)

The prefix tree in a verifiable log-backed map, which can be in the form of a \textit{sparse} Merkle tree pre-populated with all possible hashes (e.g., $2^{256}$ leaves to match all possible SHA-256 outputs)~\cite{laurie2012revocation,dahlberg2016efficient}, serves as a map (i.e., key-value store), while the history tree is used as a log that records all signed root hashes for the map, ensuring that clients can verify that the map they are shown has also been shown to others that have audited the log.
This combination of both types of Merkle trees allows for a wider range of efficient proofs than either type of Merkle tree could support on its own (i.e., append-only for the history tree, look-ups for the prefix tree)~\cite{trillianvds}.
Users, however, still need to collectively check that both Merkle trees track the same keys and values.

A third Merkle tree can be added to construct an unequivocable log derived map~\cite{andersen2019wave}, in which the first tree is a history tree log of operations, which are 
batched into a prefix tree that allows efficient lookups of operations, and the third tree records the root hashes of the second tree.

More recent work by Hu et al.~\cite{hu2021merkle} also combines history and prefix trees by proposing a history tree in which the internal nodes store the root hashes of prefix trees.
At any given epoch, the root hash of the history tree summarises the state of all prefix trees at that epoch, making it easier to monitor new changes, while the internal prefix trees make it easy to look up key values in the current epoch.
Because the history and prefix trees are part of the same tree, checking that both trees track the same keys and values is easier.

Reijsbergen et al.~\cite{reijsbergen2022tap} also combines several types of Merkle trees, this time a prefix tree in which all the leaves are the root of a Merkle sum tree in which nodes contain homomorphic commitments to the sum of the values of their child nodes, down to the \textit{value} of each leaf.
The prefix tree structure enables efficient lookups whilst the sum tree makes it possible to support a wider range of queries (sums, counts, averages, min/max, and quantiles) with integrity guarantees.



\subsubsection*{Append-only dictionaries}
Append-only dictionaries based on bilinear accumulators~\cite{tomescu2019transparency} have been proposed as an alternative to Merkle trees, enabling logarithmic-sized append-only proofs and polylogarithmic-sized lookup proofs, although high append times and memory usage, meaning this approach is not yet practical.

\subsubsection*{Blockchains}
Blockchains provide a decentralised and tamper-resistant way of updating and maintaining a global state.
Transactions that update the state are logged on the blockchain, making it possible to replay all transactions and to verify that something has happened if it is included in the blockchain, as well as when it was included.

Beginning with Bitcoin~\cite{nakamoto2008bitcoin}, blockchains have been used by cryptocurrencies to provide a transparent record of transactions over a network.
As Ethereum~\cite{wood2014ethereum} and later projects have shown, it is possible to rely on blockchains to execute arbitrary programs (smart contracts) and record these executions on the blockchain.
This allows a wide range of applications to run transparently on top of a blockchain or to use an existing blockchain to store evidence in a tamper-resistant way~\cite{goldwasser2017public,frankle2018practical,nguyen2018trusternity,panwar2019sampl,hicks2018vams}. 

Blocks in a blockchain store data (including the state of a smart contract) in Merkle trees so transparency applications that run on top of Merkle trees can be adapted to a blockchain so that its consensus protocol replaces the need for gossiping between clients that is required in a Merkle tree based system to guard against equivocation~\cite{bonneau2016ethiks,tomescu2017catena}.

Blockchains can be permissionless or permissioned.
For logging purposes, the effect of choosing one or the other is that in a permissionless setting, it is possible to use an existing public blockchain, such as Ethereum, in which case the blockchain will be maintained regardless of your use case because many other applications rely on it, as well of the value of the underlying cryptocurrency.
Thus, any incentives to maintain (or not) a reliable log are taken care of (at a price determined by the underlying cryptocurrency).

On the other hand, relevant events may not appear in an accurate chronological order because their inclusion will depend on miners who will primarily care about including the transactions that maximise their revenue rather than the needs of a single transparency application. 

The effort required to use an existing public blockchain and write a smart contract for it may also be much less than deploying an entire system like Certificate Transparency, allowing for more applications of transparency.

In a permissioned setting, known pre-determined parties will have to ensure that the log is maintained but, because there is no need for an underlying cryptocurrency, the system could be set up to include new events to the chain as they arrive rather than at the wishes of an uninterested miner.
In this case, because all parties are known and the blockchain is more likely to be application specific than a general-purpose blockchain, this setting is also much closer to deploying a Merkle tree based system like Certificate Transparency, with the benefit (or cost) of having a consensus protocol.

\subsection{Sanitisation mechanism}
The information recorded on a log will often be sensitive, in the sense that it affects the privacy of an individual or that it reveals confidential information about the system it is pulled from.
For this reason, sanitising the information that is logged will be necessary but must be done in a way that does not compromise the desired transparency.

The sanitisation mechanism determines how logged information is processed, in plaintext (i.e., \textit{unsanitised}), through a privacy-preserving form of data release (e.g., by adding noise or generating a synthetic data~\cite{hicks2018vams}), in an encrypted form to be decrypted by specific parties (e.g., designated auditors being given access to raw data, individuals accessing individual evidence~\cite{hicks2018vams}), or using cryptographic techniques such as zero-knowledge proofs to assert relevant properties of the logged information without revealing the underlying data~\cite{frankle2018practical,goldwasser2017public,panwar2019sampl}.

Access to unsanitised information may be required if no sanitisation mechanism exists that is compatible with the desired transparency.
For example, there may be no way to satisfy reasonable differentially private bounds without adding excessive noise, to produce zero-knowledge proofs that assert the necessary properties of the logged information, or simply to rely only on cryptographic proofs about data.
In such cases, it may be necessary to permit access to unsanitised data by designated auditors, while the public is given access only to sanitised data that can be used to verify the results of an audit published by the designated auditors.

Beyond the data itself, identifiers (and other metadata) that allow users to verify their individual data may also need sanitisation.
CONIKS, for example, uses a verifiable random function to produce a user identifier for the log that does not reveal the identity of the user to others~\cite{melara2015coniks}, and more recent work has introduced append-only zero-knowledge sets that minimise the leakage from queries~\cite{malvai2023parakeet,chase2019seemless}.

\subsection{Release and query mechanism}
Once data is logged, it must also be possible to release the data or perform queries on it.
As shown by Reijsbergen et al.~\cite{reijsbergen2022tap}, it is possible to implement (Merkle tree) logs in such a way that they natively support broader queries than simple lookups, but more can also be done.

Given a database, it is possible to store the hash of the database on a log, enabling users to verify that the database they are querying is the same as the one indicated by the log if they can download the entire database, but this does not guarantee the integrity of a query on that database.

Work on single client authenticated databases~\cite{zhang2015integridb,zhang2017vsql} i.e., outsourced databases that guarantee the integrity of queries and updates to the database, has led to work combining authenticated databases with a log such as a blockchain on which a smart contract is running~\cite{peng2020falcondb}.
The log ensures consistency and allows clients to verify that the database they are querying (without needing to go through the blockchain) is the database that has been recorded on the log, allowing for a broader set of queries than what is natively supported by the log itself.

Specialised formal languages, similar to TILT~\cite{grunewald2021tilt} (developed for the GDPR transparency requirements), could also be developed to produce application-specific transparency APIs that return human-readable answers to queries.

As discussed in the case of sanitisation mechanisms, data may appear in different forms to different parties e.g., only some designated auditors may be able to access raw data.
One way of doing this is simply to encrypt data under the relevant parties' public keys so that only they can decrypt the raw data, but another possibility is for the release and query mechanism to implement access control that determines who can query the log.
Depending on the type of log, this may be more or less simple. 
For example, a blockchain based system can implement access control via a smart contract.
This could also be set up to log queries if necessary.
For a Merkle tree based system the access control mechanism would have to be built on top of the logs.

\subsection{External mechanisms}

Transparency cannot be expected to be effective by itself, it must work to enable action based on what it reveals.
For example, if transparency produces evidence that a system has malfunctioned, it can allow aggrieved parties to take legal action, governance decisions about a platform or network~\cite{kiffer2017stick}, and the removal of parties from a network if they cause a fault~\cite{symantec}.
This entails supporting processes such as public discussions about the system to which transparency is applied and, for practical accountability purposes, legal processes that resolve disputes about a system or more automated processes that similarly make it possible to contest actions taken by the system.
This is a key difference between tools that evaluate the compliance of a system with preset norms e.g., the correct execution of a program, and transparency enhancing technologies that can allow the norms enforced by a program to be contested~\cite{hicks2022transparency}.

This process starts with users being able to check information that is relevant to them or being notified about such information.
Notification tools~\cite{angulo2015usable,murmann2018usable,murmann2019or,murmann2017tools,murmann2019eliciting,fischer2016transparency,fischer2013can} are a useful way to keep the user in the loop, without needing them to perform queries, when their explicit consent for an action is not required, but this does not necessarily allow a user to contest any action that is taken.

For an action to be contested, there must first be evidence of that action.
Often a program is assumed to have been correctly executed unless there is evidence of the contrary, but systems often fail to produce such evidence~\cite{murdoch2014security}.
Transparency should address this, and gossip and consensus protocols can also play a part in spreading evidence and reaching a conclusion about evidence.
What is then important is that the evidence be useful.

For an automated process, the proof must fit the requirements of the program that will evaluate it.
For a non-automated process e.g., a legal process, evidence being useful means that it should be \textit{admissible} in the relevant jurisdiction.
Admissibility involves the data itself and also the authentication of the data, its integrity, the network over which the data is exchanged, and how it is then stored~\cite{mason2021electronic}.

In both cases, this requires the form of the evidence and the process in which it will be used to be taken into account before it is produced for it to be useful.
In the non-automated case in particular, evidence is not sufficient to contest a system by itself (unlike automated processes) and the outcome of the dispute process can vary much more, up to contesting the existence and norms of the system.

In such cases, it may not always be clear when considering a single event, why the system failed~\cite{hicks2019transparency}.
This can require a broader discussion about the system and both the individual evidence and aggregate evidence (e.g., error rates) about the system to be considered to see which is more likely.
To act on information also requires the ability to understand that information, which can be made easier via explanations~\cite{rader2018explanations}, context~\cite{degbelo2018increasing}, and labels~\cite{holland2018dataset,kelley2009nutrition}.
This is particularly important, but also challenging, because disclosure practices are not always well designed~\cite{norval2022disclosure}.


\section{Transparency And Security}\label{sec:def}

Although many transparency enhancing technologies have come from security and cryptography research (e.g., cryptographic logs) and, therefore, have involved a security-focused approach, this is not always the case.
Moreover, even for cryptographic mechanisms, threats are typically expressed in terms of the cryptographic properties of the mechanisms, particularly when these mechanisms are introduced as abstract primitives, useful for applications outside of transparency, rather than as part of a system focused on transparency, which is our approach here.

\subsection{Assets and beneficiaries of transparency}
The inputs, processes, and outputs, of systems are assets for the parties that own and operate them.
The value of these assets can depend on their confidentiality.
Datasets, a codebase, a machine learning model and its outputs, can all contribute to a competitive advantage, and their confidentiality can also help avoid liability for flaws in the system, or give the illusion of technical sophistication.

Transparency can benefit system operators if it increases public trust.
This can be true regardless of whether or not the system is good by any measure because an organisation operating a flawed system may engineer a form of transparency that does not reveal these flaws by, for example, limiting transparency to only reveal favourable information.

Because transparency does not necessarily increase trust, however, operators of reliable systems may feel they have little to gain and operators of unreliable systems may have little to lose.
That is unless transparency is deployed in such a way that, for example, it harms those who operate unreliable systems by enabling consequences. 

For the public, transparency should be a valuable asset, revealing useful information about a system over which they have no control and allowing them to take action by choosing whether to use the system, contest it, and hold the system operator to account for any faults.
Privacy concerns over the public release of sensitive data that pertains to them may, however, be an important drawback.

Thus, transparency can be both beneficial and a drawback for system operators and the public, and, importantly, the ways in which the public may benefit from transparency may be a drawback for the system operator.
When this is the case, it should be ensured that blame avoidance strategies (e.g., avoidance of record keeping, gaming performance metrics) are not put into place~\cite{hood2007happens}.

\subsection{Threats based on essential mechanisms}

\subsubsection*{Logging}
The logging mechanism relates to the system operator of the system, from which information is recorded, and the log operators that maintain the log.
Assuming that the logging mechanism is based on sound cryptography (e.g., a secure hash function, public key encryption scheme, and digital signature scheme) then what remains as a threat is the ability of a malicious system operator (or whichever party is responsible for logging information) that attempts to compromise what makes it to the log in the first place.

\subsubsection*{Sanitisation}
As sanitisation can take place before or after information is logged, threats can come from either the system operator (before logging) or from data releases and queries (after logging).

A system operator could try to compromise a sanitisation mechanism just as they would the logging mechanism itself.
A sanitisation step taking place before the information is committed to the log would be intended to work towards the confidentiality of commercially sensitive information about the system or to respect the privacy of users who relate to logged data.
This could be abused by the system operator to hide other information without having to compromise the logging mechanism.

For a sanitisation mechanism that takes place after information is logged, threats are posed by parties attempting to learn private information about others from the information they have access to.

The sanitisation mechanism could also be used by log operators, if sanitisation is done at the interface between the log and users of transparency, or auditors, if they are given access to raw information that they sanitise for public release, to compromise the information that is released. 
This can be achieved either by producing sanitised information that does not relate to the original information (e.g., releasing wrong statistics) or relying on an honest use of a sanitisation mechanism that obfuscates some information as part of its use e.g., by adding noise.

\subsubsection*{Release and query}
The form of the information made available by release and query mechanisms will depend on the sanitisation mechanism, so the threats that are specific to release and query mechanisms will be those that target the access control it implements and the integrity of the information (sanitised or unsanitised) that is released.
Given that information should broadly be released to everyone except for individual evidence (available only to data subjects) and unsanitised information (available only to trusted auditors), the threat is that any other party may try to pose as an individual or trusted auditor to gain access to their privileged information.
The right to access under the GDPR has been abused for this purpose~\cite{di2019personal}, as well as to infer information about the organisation answering the query ~\cite{singh2019security}.

If information is simply released, without the need for queries, threats could be posed by having only a partial release of information, or a different release of information to different users.
When queries are involved, the threats are that the query mechanism could constrain acceptable queries to queries that are not practically useful. 
It could even do so for a priori valid reasons such as limiting the privacy loss associated with queries, as in a differential private query model once the privacy budget is used up.
A limited query mechanism could also serve to require an impractically large number of queries to obtain any useful information.

\subsubsection*{External}
External mechanisms (not necessarily technical mechanisms) represent the interactions between users of transparency and the actions that they can take based on it.
The threat in this case is misinformation and disinformation and the threat actor can be any user of transparency giving (mistakenly or intentionally) inaccurate information.

This can be seen as an attack on the integrity of the information made available through transparency, which can be mitigated by making information verifiable i.e., ensuring that the same information (barring individual evidence) is available to all.
In the specific case of individual evidence, it should be ensured that an individual cannot lie about their individual evidence, but also that they can use that to show that any individual evidence they disclose is correct.

\begin{table*}[t]
\begin{center}
\caption{Threats for transparency enhancing technologies based on editorial control (EC) and individual evidence (IE).}
\label{table:threatsummary}
\resizebox{\linewidth}{!} {%
\begin{tabular}{l l l l }
\textbf{Mechanism} & \textbf{Threat} & \textbf{Affected transparency property} & \textbf{Threat actor(s)} \\
\midrule
Logging & Compromised logging mechanism (EC, IE) & Integrity  & System operator \\
& Compromised log server (EC, IE) & Integrity, Availability  & Log operator \\
& Collusion between system operator and log operators (EC, IE)& Integrity, Availability & System operators, log operators \\
\midrule
Sanitisation & Loss of privacy for data subjects & Respect of privacy and confidentiality & Users of transparency \\
& Control over logging (EC. IE) & Availability & System operator \\
& Control over release and query responses (EC, IE) & Integrity & Log operators, auditors \\
\midrule
Release \& Query & Access to raw data or individual evidence & Respect of privacy and confidentiality & Users of transparency\\
& Restricted releases (EC, IE) & Availability, interpretability & Log operators, auditors \\
& Constraints on queries (EC, IE) & Availability, interpretability &  Log operators\\
\midrule
External mechanisms & Misinformation \& disinformation & Interpretability & Auditors, data subjects, third parties \\
& Lying about individual evidence  (IE) & Trustworthiness & Data subjects \\
& Discrediting individual evidence (IE) & Actionability & Third party individuals \\
\hline
\end{tabular}
}
\end{center}
\end{table*}

\subsubsection*{Editorial control and individual evidence}
Examining different attempts to implement transparency around the world, Taylor and Kelsey found that the two general threats to transparency were editorial control i.e., the ability to control what is made transparent, and individual evidence i.e., the ability to suppress the ability of a person to find information that relates to themselves through transparency~\cite{taylor2016transparency}.

We relate this to the mechanism-specific threats we have outlined above in Table~\ref{table:threatsummary}.
Both editorial control and lack of available individual evidence can occur through the system operator (logging mechanism), and the log operators and auditors (sanitisation and release and query mechanism), resulting in effects on the external mechanisms.
\section{Transparency Infrastructure}\label{sec:infra}

\subsection{Requiring and maintaining transparency}
Deploying transparency requires an infrastructure that supports the operation of logs and the storage of any data required, including data that may not be stored on the log.
Because logs (and any other data) may be used after the system (or its operator) it originates from stops operating, they must be stored independently from the system.
Thus, although a centralised approach could be sensible on the basis that only the system operator has a business reason to store that information, it may not be reliable for transparency.

Relying on distributed storage, however, raises questions about how to distribute it.
Parties such as NGOs monitoring government activities or public institutions monitoring some businesses may have a strong incentive to support transparency infrastructure that relates to issues that they investigate as it directly supports their goals.

This can also be the case in commercial settings.
Google, for example, is responsible for the design and deployment of Certificate Transparency.
Because Google Chrome is the dominant browser~\cite{browsershare}, it has a direct interest in keeping Certificate Transparency operational, requires that any certificate appears in at least two logs, and operates some of the logs itself.
(Google previously required one of the two logs to be a log operated by Google~\cite{meiklejohn2022sok}.)

Unfortunately, this example does not generalise well.
In most cases, the parties that design the transparency enhancing technology may not be those that operate it, or may not have a direct incentive to ensure its success or the resources both in terms of influence on the ecosystem and technical resources (e.g., in the case of NGOs) to guarantee it.
Proponents of blockchains and cryptocurrencies argue that they offer the possibility of designing decentralised systems that, via mechanism design, can ensure that participants in the system have incentives -- typically financial -- that are aligned with maintaining the system.
Blockchain can then serve as logs, requiring only a smart contract to deploy, and services such as Filecoin~\cite{filecoin} could also offer decentralised storage when it is necessary to store more than logs.

Users themselves could drive businesses to provide greater transparency as they do react to, for example, being shown the extent to which they are tracked~\cite{weinshel2019oh} and how moderation is applied~\cite{jhaver2019does}.
However, they often have to rely on tools set up by system operators that do not provide complete transparency, or transparency that users can understand~\cite{andreou2018investigating,urban2019your}.
As we already noted in Section~\ref{sec:def}, system operators may not be incentivised to provide effective transparency, leading to a market for lemons.

Regulation could also play a part by imposing a statutory requirement to provide transparency could be through enforcement action of a regulator such as the Federal Trade Commission or a data protection authority.
The European GDPR, which effectively applies globally to any service that has users who are citizens of the EU, notably includes several articles concerning transparency.

Designated auditors may also have the power to ask for the infrastructure needed to operate a transparency enhancing technology.
For example, the IPCO in the UK is tasked with auditing how law enforcement access telecommunications data (a yearly report is published~\cite{ipco2020}) and can require that public authorities and telecommunication operators provide any assistance required to carry out audits, which could include implementing IT infrastructure~\cite[Section 235(2)]{IPAct}.

Some regulations e.g., the German Network Enforcement Act (NetzDG), do include require transparency requirements about, for example, how unlawful content is dealt with and have resulted in fines for companies such as Meta. 
Companies differ in how they implement their compliance with this regulation~\cite{wagner2020regulating} and are likely to differ in implementing any other kind of transparency requirement.
Standardisation may, therefore, be required if there is any hope of achieving reliable transparency across different types of systems, and this should be done taking into account threat models and mechanisms to deal with these threat models, and still allow enough flexibility to adapt to, for example, case-specific sanitisation needs.

In particular, because regulators are not the people affected by flawed systems and can typically only levy fines on system operators who treat these as a cost of business, transparency that provides information to regulators is unlikely to offer much progress.
Transparency that is user-facing, and can inform users in a way that allows them to take action on the basis of that information may be more effective.

\subsection{Truth}
A limitation of logs is that their security properties cannot ensure that any logged data or event is true.
Dealing with this depends on how the logging mechanism can ensure that the recorded value matches that of the object of interest, and what the logging mechanism actually records.

In Bitcoin, miners reach consensus on which public keys own each bitcoin.
A user may want to send bitcoins to another user but if the transaction is dropped by the network the transaction fee was too low, then the transaction is never executed or recorded.
Thus, the Bitcoin network is transparent about how the miners view the network, not about every action of the users in the network.

Moreover, not all real-world transactions are logged because Bitcoin private keys may be exchanged offline with no mapping between keys and identities to restrict this. 

Likewise, Certificate Transparency is transparent with respect to the set of certificates accepted by log servers, not with respect to all certificates emitted by certificate authorities as some may not be logged.
Browsers can reject certificates that do not appear in Certificate Transparency logs, however, which ensures that log servers that are operated by, for example, Google, have the incentive to log all valid certificates sent to them by certificate authorities.

The interface between the device that records information that is logged and the log is also important.

A malicious recording device would be a clear weakness so a trusted hardware interface could be used.
The security of trusted hardware components may, however, be centralised if all units are the same.
If one unit is broken then, for example, the attestation key could leak~\cite{van2018foreshadow}, rendering all other units worthless.
This is a case of weakest-link security that depends on the party with the lowest benefit-cost ratio in securing their unit~\cite{varian2004system}, in a scenario where that party may be adversarial and have full physical access to their hardware.

Alternatively, it may be possible to rely on non-colluding parties to cross-verify information.

Problems may also occur if there is no ground truth for the logged data.
For example, wage transparency could identify wage gaps but if the party that logs salaries is the business itself, the logging mechanism (or any computation used to identify a wage gap~\cite{lapets2018accessible}) can execute correctly regardless of the data (and the resulting analysis) being true if individuals cannot verify their inclusion in the computation.

Problems can also occur when dealing with physical objects e.g., paper documents, because this requires a secure way of mapping physical objects to digital objects that can be authenticated once logged.
Mechanisms that provide cryptographic-like mechanisms to authenticate certain physical objects do exist, however.
There is a body of work that studies how paper documents could be authenticated based on their physical characteristics~\cite{toreini2017texture,sharma2011paperspeckle,clarkson2009fingerprinting,buchanan2005fingerprinting,samsul2010recognizing,li2018printracker,guarnera2019new,van2006recognition,9381886}.
This would allow the document to be logged with its fingerprint, allowing it to be authenticated later if required.

\section{Balancing Transparency With Privacy}\label{sec:privacy}

Because privacy concerns can create legitimate restrictions on transparency, privacy enhancing technologies that preserve privacy while retaining the utility of information can enable transparency.
(In turn, transparency can help users identify privacy risks~\cite{henze2016towards,van2017better,li2017privacystreams,de2018privacy}.)

There are two types of information to consider, aggregate information related to a population and information related to individuals.
Aggregate information makes it possible to determine how the system is functioning as a whole e.g., whether it is (un)fair, (un)biased, or error-prone.
By itself, this can be enough to reach a conclusion about the system e.g., whether the system should be modified, shut down, or to make the choice of participating in the system. 
For individuals, it is also important to be able to determine how they are personally affected by the system as, for example, a biased system will not impact all users in the same way.


In the case of aggregate information, the privacy requirement is that the aggregate information should not leak information about an individual, including the inclusion of an individual's data in the data that was used to produce aggregate information.
This often involves differentially private mechanisms that determine the kind of perturbed data that can satisfy data protection requirements~\cite{cohen2020towards,nissim2017bridging}, and zero-knowledge proofs, which allow the execution of a process to be verified without revealing anything else about the process~\cite{bamberger2022verification,goldwasser1989knowledge,goldreich1986prove}.


For individual information, controlling access to information also matters since revealing information only causes a loss of privacy if it is revealed to someone other than the individual it relates to. 

While differentially private mechanisms and zero-knowledge proofs appear necessary to balance transparency and privacy requirements, there are concerns tied to editorial control and individual evidence that we consider here.

\subsection{Editorial control}
Editorial control encompasses not only the ability to prevent access to information (e.g., information being logged by the transparency enhancing technology) but also any way of influencing what is or is not recorded, the format in which it is recorded, what is shared with who, and the terms under which information is shared.

Differential privacy does this by changing the information that is shared, for example through the addition of noise or by sharing a synthetic dataset rather than the original one.
While differentially private mechanisms work to preserve as much utility as possible, this is nonetheless a form of editorial control that can work in favour of an adversarial system operator.
This is because the addition of noise disproportionately affects less represented groups in the data.
For example, the adoption of differential privacy for the U.S. Census could effectively erase smaller towns from census data~\cite{nyt}.
More generally, differential privacy could be used, under the cover of it being a required privacy enhancement, as a way of masking bad outcomes on minority groups, or to make low-frequency faults disappear.

Another way in which differential privacy can lead to editorial control is by limiting the number or type of queries that can be made as part of the query mechanism of the transparency enhancing technology.
Differential privacy assigns a privacy budget that dictates how many queries can be made (based on their sensitivity), placing a limit on what and how much data subjects, third-party auditors, and third-party individuals can do through a query mechanism.
It could also allow an adversarial auditor (perhaps colluding with the system operator wanting to work against transparency) to exhaust the privacy budget by performing high-sensitivity queries that do not reveal anything unwanted.

However, this can be avoided by relying on a release mechanism that generates synthetic data (although not a general solution~\cite{stadler2021synthetic}) that can be queried ad infinitum, rather than relying on a query mechanism that serves differential private answers to queries on the database of original data.

Zero-knowledge proofs can also act as a form of editorial control.
A zero-knowledge proof reveals nothing but the truth of a statement, which can remove context from a query.
Requiring that any query by an auditor be expressed as a provably true or false statement within the constraints of a formal language may also restrict the range of possible queries, and prevent necessarily vague queries.

Querying for provable statements can also be made inefficient this way e.g., iterating over queries of the type ``is the number of data points with attribute $\alpha$ greater than $x$'', as queries must be designed without access to data. 
The result of this is that practically speaking, it is only possible for auditors and individuals to verify statements that are given to them by those who control the information that is queried, rather than being able to perform their own investigation.

Moreover, detecting a flawed implementation of a zero-knowledge proof system that allows counterfeit proofs to be produced can be hard.
Flaws in zero-knowledge proof systems have only happened by accident so far~\cite{miller2022coordinated,zcashbug}, but there is a precedent for cryptosystems that could plausibly be exploitable by design~\cite{bernstein2016dual}.
A malicious system operator could attempt to introduce an intentionally flawed zero-knowledge proof system that would allow them to appear compliant with any desired norm.


\subsection{Individual evidence}
Individual evidence is desirable for the simple reason that a general overview of a system may reveal issues with the system e.g., it is biased against certain attributes or has bugs, but fail to show their impact on individuals e.g., whether one was discriminated against or affected by a bug.
This requires not only knowledge of the system's outcome for that individual, which usually will be known for the outcome to have any effect although this may not always be the case (e.g., for confidential processes) but also some form of ground truth for what the outcome could have been, which in general may be harder to obtain.

For example, the covid-19 pandemic caused secondary education exams in the UK to be cancelled in 2020 and grades to be awarded based on an algorithm using results of past students as input.
The population outcome was normal by design -- the distribution of grades matched historical distributions for each school -- but it meant that students who performed outside the historical norm could be awarded lower or higher grades than expected for the sake of preserving the historical grade distribution.
Individual evidence in the form of teacher predicted-grades, however, made it possible to easily identify how students had been affected (e.g., a student with a high teacher-predicted grade being awarded a low grade) and the algorithmic marking scheme was quickly replaced with teacher-predicted grades~\cite{bbc}.


Individual evidence can also be useful when there is a dispute about whether an individual has made an error when using a system or has been a victim of a bug.
Human errors and bugs can happen at reasonably low frequencies so conclusively determining whether one is more likely than the other can be impractical, and neither the presence of bugs in the system nor the possibility of a human error can be used to invalidate the other~\cite{hicks2019transparency}.
Individual evidence that makes it possible to identify the error in an event log and a record of actions by the individual could make it much more efficient to determine whether the error was human or due to a bug in the system.

The role of privacy enhancing technologies, however, is often to make it impossible to link an individual to an input or output of the subject system's process.

Differential privacy guarantees that an individual does not have too much of an effect on outputs so that it cannot be determined their data was used to obtain that output without an additional mechanism that deals with this.

Zero-knowledge proofs remove the relation between the output of the computation it verifies and its inputs. 
If individual evidence exists, however, a zero-knowledge proof could be used to show an individual that their individual evidence was used in the computation.
Without this, an adversarial system operator or auditor could simply use inputs that they choose or generate to obtain valid zero-knowledge proofs for whatever they want.


This means that the use of these privacy enhancing technologies to allow the release of aggregate information requires that additional mechanisms be used for individuals to obtain the individual evidence necessary to contextualise the aggregate information and the effect the system has had on them.

\section{Case Studies}\label{sec:cases}

\subsection{Certificate Transparency}
SSL certificates are an essential part of web security, allowing a user's browser to verify the owner of a website.
Certificates are issued and signed by trusted third parties, certificate authorities, who can be the source of security incidents~\cite{durumeric2013analysis,clark2013sok}
An example of this is the DigiNotar hack~\cite{van2013diginotar}, which led to hundreds of rogue certificates being issued with DigiNotar's signing key and DigiNotar certificates being rejected by most browsers~\cite{microsoft,google,firefox}.

Certificate Transparency~\cite{laurie2014certificate,stradling2021certificate} was developed to address this type of incident.
Acknowledging that it is not possible to prevent rogue certificates from being issued, Certificate Transparency works by making certificate issuance transparent and working against malicious certificate issuance by helping reveal cases where this happens.
This is achieved by using logs based on Merkle history trees that ensure the list of logged certificates is a secure append-only transparency overlay~\cite{dowling2016secure,chase2016transparency}.

Certificate authorities submit certificates to the logs themselves and browsers will only accept certificates that come with a signed certificate timestamp from log servers, so a malicious certificate authority cannot compromise the efficacy of the logging mechanism by not submitting certificates that they issue to logs and collusion between a certificate authority and a log server is mitigated by requiring multiple signed certificate timestamps from different logs.

Certificate Transparency is widely deployed, with the percentage of main-frame HTTPS page loads and HTTPS connections with at least two valid signed certificate timestamps reaching above 60\% as of 2018 for Chrome users~\cite{8835212}.
There is significant infrastructural backing from organisations like Google, Mozilla, and Cloudflare, and free services such as Let's Encrypt~\cite{aas2019let}.


There is no sanitisation mechanism involved in Certificate Transparency, although some interactions involve privacy concerns for users.
For example, when their browser queries a proof of inclusion in a log, it reveals the website that the user is browsing. 
As a result, most clients do not directly request proofs of inclusion, although solutions based on fuzzy ranges, private set intersection, and private set membership protocols have been proposed~\cite{meiklejohn2022sok}.

Reporting that a certificate has not been included in a log also reveals a user's browsing activity for that website.
This can be mitigated by using zero-knowledge proofs to allow the browser to prove to a browser vendor (e.g., Google) that it knows a signed certificate timestamp signed by a log server (without revealing it) despite the log omitting this certificate, therefore showing that the log does not have integrity~\cite{eskandarian2017certificate}.
This approach has downsides, however, as it would require changes to log implementations and APIs, and obfuscate details in investigations of log misbehaviour~\cite{stark2021certificate}, showing the tension between transparency and privacy goals.

Other issues exist with the certificates themselves and logs, which can be used to identify potentially vulnerable websites because websites with expired certificates tend to more outdated software that may be vulnerable to CVEs~\cite{pletinckx2023certifiably}.
The volume of information available through Certificate Transparency also makes it possible to monitor logs to identify new DNS names i.e., service endpoints,  that may be vulnerable to an attack, rather than inefficiently scanning the IP space~\cite{scheitle2018rise}.
Logs can also be mined to detect subdomains, as well as other sensitive information including names, usernames, email addresses, business relationships, and unreleased products~\cite{roberts2019certificate}.

The volume of logged certificates poses scalability issues as well.
Monitors, who fetch and try to spot suspicious certificates, cannot guarantee that fetching certificates returns a complete set of certificates, meaning that fraudulent certificates may be logged but not spotted~\cite{li2019certificate}.



External mechanisms play an important role in Certificate Transparency.
Certificates must be revoked as time passes or in the event of an incident (e.g., DigiNotar).
In such a case, a human decision must be made based on the information available and the potential to act on that information.
The latter means that power is concentrated in browser vendors (e.g., Google, Mozilla, Microsoft, Apple, Brave) which are the only parties who can act on certificate transparency revealing a malicious or compromised certificate authority by blocklisting it.
Expert users can in principle also inspect logs, but represent a tiny minority of users.

Gossip protocols should play a role in enabling clients to exchange messages containing warnings or inconsistencies between signed tree heads of logs~\cite{chuat2015efficient}, but gossiping is not widespread~\cite{gasser2018log}.
There are several ways to work around this, replacing gossiping as a type of external mechanism with a protocol that is integral to the transparency overlay.

The first way is to use a blockchain and rely on its consensus protocol for consistency~\cite{bonneau2016ethiks,tomescu2017catena}, but this can be expensive because of transaction costs and has slow finality if relying on a slow blockchain (e.g., Bitcoin or Ethereum).

The second way is to rely on witnesses (e.g., the different Certificate Transparency log servers) could collectively sign a checkpoint of a log, producing some form of consensus that the log has been verified up until the checkpoint~\cite{meiklejohn2020think}, but this could suffer from liveness issues if there are too few witnesses.

\subsection{Blockchain based cryptocurrencies}
Cryptocurrencies, such as Bitcoin~\cite{nakamoto2008bitcoin}, Ethereum~\cite{wood2014ethereum}, and many others, aim to enable decentralised peer-to-peer transactions between users that do not rely on any centralised institutions such as banks, Paypal, and VisaNet~\cite{nakamoto2008bitcoin}.

This requires solving the problem of currency minting and double-spending such that no single user can unilaterally determine the amount of tokens they control, or spend the same tokens multiple times.
This is achieved by relying on a blockchain, which records blocks of transactions (that refer to the previous block in the chain), which are mined (i.e., validated) by miners expending a scarce resource such as computational work (e.g., proof-of-work, proof-of-storage) or stake in the currency (proof-of-stake) for the right to mine blocks.
The state of the blockchain is public and agreed upon by the nodes in the network through a consensus protocol, allowing anyone to track any asset on the network.

Chase and Meiklejohn~\cite{chase2016transparency} considered the Bitcoin blockchain as one of their two case studies (the other being Certificate Transparency) in their formalisation of transparency overlays.
The important difference between the two systems that emerged is that miners in permissionless blockchain systems are not known and, therefore, cannot be held responsible for faults and are not trusted to provide consistent views of the blockchain.
This can be dealt with through penalties and \textit{slashing} mechanisms that exist in proof-of-stake cryptocurrencies, such as Ethereum~\cite{slashing}, to directly fine or remove from the network block validators that misbehave because being elected to be a block proposer or validator requires staking funds.

Nonetheless, although it is possible to see what is going on with blockchain explorers (e.g., \url{https://www.blockchain.com/explorer}) that display the latest block information, users must download, store, and verify the entire blockchain to assure themselves they have the correct information.

As blockchains record an increasing number of transactions they become larger and more expensive to download, store, and verify.
For example, the Bitcoin and Ethereum blockchains now amount to hundreds of gigabytes of data, making it difficult for most users to operate a node that independently verifies the state of the blockchain.
As a result, users often run light clients that verify only block headers and the transactions inside blocks, decreasing security.

Transparency in this setting, whether at the stage of validating blocks or later auditing past transactions, is useless if it is not used to verify the system's consistency and ensure that only valid transactions are processed, so this is a problem that relates to the transparency of the system.

One approach to solving this issue is based on succinct blockchains that reduce the computational costs of verifying the blockchain~\cite{kiayias2020non,bonneau2020mina}.
Recursive succinct arguments of knowledge can be produced in time proportional only to the number of transactions added since the previous block and verified in constant time~\cite{bonneau2020mina}.
To verify the blockchain, this allows blockchains to effectively be compressed from hundreds of gigabytes (the size of a blockchain after a few years) to a 22 kilobyte proof that verifies transactions and consensus rules, which can be verified in milliseconds.

Another approach is based on fraud proofs, which involve full nodes producing proofs of invalid transactions that light clients can efficiently verify to narrow the security gap between full nodes and light clients~\cite{al2021fraud,yu2020coded}.
Fraud proofs also play a role in enabling scaling solutions such as optimistic rollups on Ethereum~\cite{optimistic}, which process transactions off the main chain (reducing congestion and transaction fees) and then post only compressed transaction data on the main chain.
The transparency obtained from the transaction data posted on the main chain makes it possible to verify the validity of transactions and produce fraud proofs for any invalid transactions. (Zero-knowledge rollups, the alternative to optimistic rollups, rely instead on proofs of validity to prevent invalid transactions~\cite{zkrollup}.)

Another commonality with Certificate Transparency is that blockchains do not necessarily offer much in terms of sanitisation mechanisms, and there is no right level of privacy that is agreed upon, between full transparency that compromises basic privacy expectations and fully obfuscated transactions that rely on the blockchain as an integrity check rather than a transparency mechanism.

Early systems, such as Bitcoin and Ethereum, do not offer any privacy because, although they are pseudonymous, it is easy enough to identify unique users by studying the public transaction flows recorded on the blockchain~\cite{meiklejohn2013fistful} and trace coins that have been used as part of some unwanted activity~\cite{anderson2019bitcoin,ahmed2018tendrils}, a practice that has been commercialised by companies such as Chainalysis, TRM, and Elliptic.

More recent systems have attempted to provide greater privacy~\cite{almashaqbeh2021sok} through the use of zero-knowledge proofs (e.g., Zcash~\cite{sasson2014zerocash}), ring signatures (e.g., Monero~\cite{alonso2020zero}), coin mixing services (e.g., Tornado cash~\cite{pertsev2019tornado}, sanctioned by the US Treasure since August 2022~\cite{sanction}), and network level mixing (e.g., Nym~\cite{diaz2021nym}).
Not all attempts have been successful in achieving their privacy goals because of low adoption, design flaws, and the inherent availability of auxiliary information available via blockchain analysis that can be exploited~\cite{kappos2018empirical,biryukov2019privacy,biryukov2019privacyb,Moser,hong2018practical,wu2022tutela}.

Balancing privacy goals with the goal of stopping tainted funds (e.g., stolen funds) from being laundered through, for example, mixing services has also been shown to be possible.
One possible solution is to produce a zero-knowledge proof that the funds one has put through the mixing service did not come from any address that is publicly associated with tainted funds.
In this case, the transparency that allows the addresses containing stolen funds to be identified would allow other addresses to use privacy services without the risk of facilitating the laundering of stolen funds~\cite{privpools}.

Another possible solution is collaborative deanonymisation~\cite{keller2021collaborative}, which would allow users to contribute information that helps identify a source of coins processed by a mixing service, enabling transparency that can be determined by users themselves rather than system designers.




External mechanisms also play an important role in blockchains and their governance.
The blockchain can show miner behaviour such as front-running~\cite{eskandari2019sok}, evidence of hacks, trace stolen funds, and so on.
This has led to important debates about, for example, whether the 2016 DAO hack on Ethereum should be reversed with a hard fork (leading to the split between Ethereum and Ethereum Classic)~\cite{kiffer2017stick}, or whether the size of Bitcoin blocks should be increased (leading to Bitcoin Cash and Bitcoin SV).

Social influence also plays a role in such discussions as public figures (e.g., Vitalik Buterin for Ethereum) and influential companies (e.g., Blockstream employed many Bitcoin Core developers) can sway public opinion.
In principle, anyone can suggest improvements and fork a blockchain to implement their suggested improvements and publicly showcase them.
Thus, although miners have the power to enforce changes as they run the software and validate transactions, and the few developers with write access to the software repositories have privilege over the code, transparency enables some redistribution of power as discussions can be based on entirely public information.

\section{Related Work}\label{sec:related}
A number of past surveys related to transparency enhancing technologies exist.
Murmann and Fischer-H{\"u}bner~\cite{murmann2017tools} focus on the usability of transparency enhancing technologies.
Hedbom~\cite{hedbom2008survey}, Janic et al.~\cite{janic2013transparency}, and Zimmermann~\cite{zimmermann2015categorization} focus on transparency tools that can be used to help users control or verify their privacy online.
Spagnuelo et al.~\cite{spagnuelo2019accomplishing,spagnuelo2019transparency} look at transparency enhancing technologies in the context of providing and complying with the transparency required by the GDPR.

In contrast to these papers, our focus is not specifically on existing tools (although we survey some and consider two use cases), but more generally on how to design and build transparency enhancing technologies based on cryptographic logs under realistic threat models that consider issues of editorial control and access to individual evidence.

\section{Conclusion}\label{sec:conclusion}
This paper provides a systematisation of log based transparency enhancing technologies, identifying the requirements and essential mechanisms of transparency enhancing technologies, and showing how threat models relate to issues of editorial control and individual evidence.
There are many use cases for transparency: Certificate and Key Transparency~\cite{stradling2021certificate,laurie2014certificate,keytransparency,melara2015coniks,bonneau2016ethiks,nguyen2018trusternity,chase2019seemless,malvai2023parakeet}, cryptocurrencies~\cite{nakamoto2008bitcoin,wood2014ethereum,sasson2014zerocash,alonso2020zero}, binary transparency~\cite{nikitin2017chainiac,al2018contour}, decentralised authorisation~\cite{andersen2019wave}, and socially driven applications such as transparency about wage gaps~\cite{lapets2018accessible}, financial markets~\cite{flood2013cryptography}, legal processes~\cite{goldwasser2017public,frankle2018practical,panwar2019sampl}, data sharing~\cite{hicks2018vams} and usage~\cite{seneviratne2014enabling}, data mining~\cite{weitzner2006transparent}, inference~\cite{weitzner2006transparentt}, advertising~\cite{venkatadri2018treads}, and open government data~\cite{shadbolt2012linked,o2011transparent}.
Many of these rely (or could as they adapt their threat models) on logs and sanitisation mechanisms as we have described.

There are clear challenges to tackle, relating to the infrastructure that would enable transparency, and balancing transparency with privacy and confidentiality concerns.
The two case studies we have provided, Certificate Transparency and cryptocurrencies, show how many of these challenges arise in practice for each essential mechanism and, in some cases, how they can be addressed.

Several additional challenges must also be resolved for transparency enhancing technologies to be practically useful in supporting users and processes such as legal disputes, in which they will engage based on what transparency reveals, and regulations that require transparency.

As we have discussed, there are many possible use cases and approaches that can be taken in designing and deploying transparency enhancing technologies.
Based on the history of transparency, effectiveness is not guaranteed.
The design of transparency enhancing technologies should, therefore, ensure that any technological attempt to enable greater transparency focus on making transparency not a goal in itself but a tool that serves a broader aim in the system in which it is put in place.
We hope that this paper supports this.


\bibliographystyle{plain}
\bibliography{refs}

\end{document}